\documentclass[aps, prl,reprint,superscriptaddress,bibnotes]{revtex4-1} 

\usepackage[T1]{fontenc}
\usepackage[utf8]{inputenc} 
\usepackage[australian]{babel}
\usepackage{svg}

\usepackage[a4paper,centering,hmargin=1.75cm,vmargin=2cm]{geometry} 
\usepackage{amsmath,amssymb,graphicx,bm,microtype} 
\usepackage[colorlinks,allcolors=blue!50!black]{hyperref} 
\usepackage[all]{hypcap}
\usepackage{cleveref,braket,siunitx}
\usepackage[version=4]{mhchem}
\usepackage{lipsum}
\usepackage{braket}

\usepackage{verbatim}

\begin{document}


\title{Reconstructing the phonon distribution of trapped ions out of the Lamb-Dicke regime} 

\author{Christophe~H.~Valahu}
\thanks{These authors contributed equally to this work.}
\email{christophe.valahu@sydney.edu.au}
\affiliation{School of Physics, University of Sydney, NSW 2006, Australia}
\affiliation{Sydney Nano Institute, University of Sydney, NSW 2006, Australia}

\author{Prachi~Nagpal}
\thanks{These authors contributed equally to this work.}
\affiliation{School of Physics, University of Sydney, NSW 2006, Australia}

\author{Teerawat~Chalermpusitarak}
\affiliation{School of Physics, University of Sydney, NSW 2006, Australia}
\affiliation{Sydney Nano Institute, University of Sydney, NSW 2006, Australia}

\author{Maverick~J.~Millican}
\affiliation{School of Physics, University of Sydney, NSW 2006, Australia}

\author{Cameron~McGarry}
\affiliation{School of Physics, University of Sydney, NSW 2006, Australia}
\affiliation{Sydney Nano Institute, University of Sydney, NSW 2006, Australia}

\author{Frank~Scuccimarra}
\affiliation{School of Physics, University of Sydney, NSW 2006, Australia}

\author{Vassili~G.~Matsos}
\affiliation{School of Physics, University of Sydney, NSW 2006, Australia}

\author{Hon-Kwan~Chan}
\affiliation{School of Physics, University of Sydney, NSW 2006, Australia}

\author{Ting~Rei~Tan}
\email{tingrei.tan@sydney.edu.au}
\affiliation{School of Physics, University of Sydney, NSW 2006, Australia}
\affiliation{Sydney Nano Institute, University of Sydney, NSW 2006, Australia}

\begin{abstract}
Characterizing the phonon distribution of trapped-ion motional states is essential for many applications in quantum technologies, but existing methods become ill-conditioned beyond the Lamb-Dicke regime. We overcome this limitation by recasting phonon-state reconstruction as a filter-function inversion problem. Using composite pulses on the carrier and multiple sidebands, we engineer well-conditioned filters in phonon space to characterize pure and mixed states, study heating dynamics, and reconstruct Fock states up to $N=250$.
\end{abstract}

\maketitle


The secular vibrational modes of trapped ions are near-ideal quantum harmonic oscillators, providing a fundamental platform for studying quantum mechanics~\cite{Meekhof1996, Leibfried2003}. Beyond their intrinsic interest, these oscillators are also powerful resources for quantum technologies: they mediate multi-ion entangling gates~\cite{Cirac1995, Srensen1999}, enable hardware-efficient quantum error correction~\cite{Fluhmann2019, Matsos2025}, provide a versatile resource for quantum simulation~\cite{Valahu2023, Whitlow2023, Navickas2025, So2025, Sun2025, McGarry2026}, and allow quantum-enhanced metrology beyond classical limits~\cite{McCormick2019, Gilmore2021, Valahu2025}. In many such cases, the capabilities provided by oscillators are enhanced by increasing the number of phonons in the motional state~\cite{McCormick2019, Wolf2019, Zhang2018}. Accurate characterization of trapped-ion motional states therefore requires methods that remain reliable over a broad range of phonon numbers, enabling both fundamental studies and the use of these oscillators as quantum resources.

Existing characterization tools use spin-motion interactions to infer information about the oscillator, including mean occupation number~\cite{Monroe1995}, phonon distribution~\cite{Meekhof1996, Ding2017b,Rasmusson2021,Mallweger2023}, density matrix~\cite{Leibfried1996, Jia2022, Simoni2025}, and phase-space quasiprobability distribution~\cite{Ding2017a, Flhmann2020, Jeon2025}. However, extending these approaches to characterize arbitrary states over arbitrary phonon ranges remains an outstanding problem. Indeed, at high phonon numbers, the breakdown of the Lamb-Dicke (LD) approximation introduces strong nonlinearities in the spin-motion interaction, posing three challenges for reliable characterization. First, accurately reconstructing an arbitrary state over a broad phonon-number range requires a correspondingly large number of independent measurements to uniquely determine the state. Second, spin-motion couplings depend nonlinearly on phonon number and can become degenerate, making different occupations difficult to distinguish. Third, even when the measurements are sufficient to distinguish different occupations, the restricted set of experimentally accessible spin-motion interactions can lead to poorly conditioned inference.

Here, we introduce and experimentally demonstrate a phonon-number reconstruction protocol that operates beyond the LD regime and is applicable to arbitrary states over arbitrary phonon ranges. The key idea is to adopt a filter-function framework to guide the measurement design. We combine measurements from multiple spin-motion interactions to ensure occupation distinguishability across broad phonon ranges, and use composite-pulse sequences to improve the conditioning of the reconstruction. We experimentally reconstruct the phonon distributions of Fock states well beyond the LD regime with occupations up to $N=250$, as well as coherent and thermal states. These capabilities enable the direct observation of highly excited motional dynamics, which we demonstrate by measuring the heating dynamics of an $N=100$ Fock state. 

\begin{figure*}[ht!]
    \centering
    \includegraphics[]{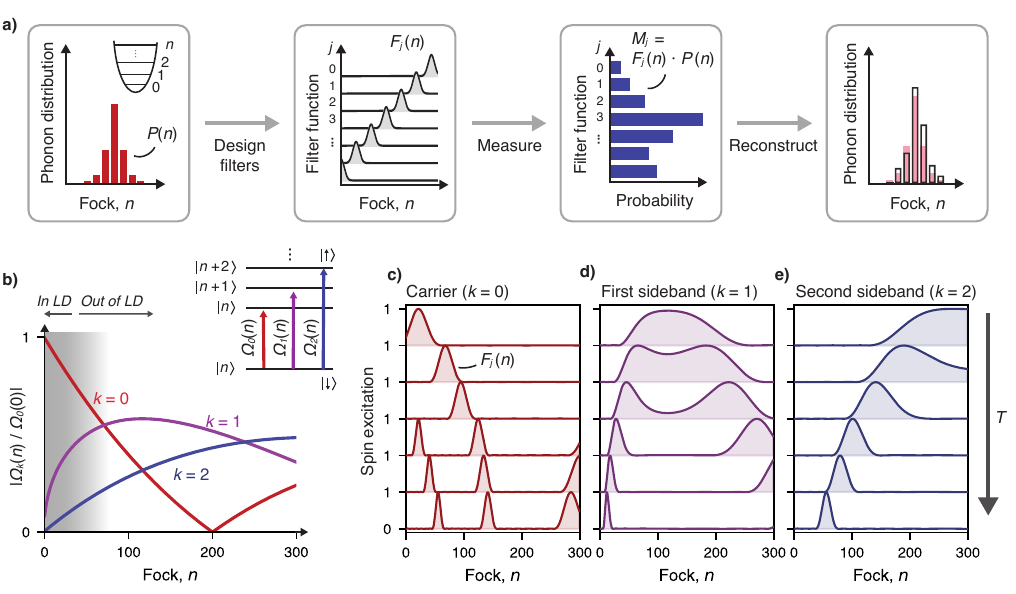}
    \caption{Reconstructing the phonon distribution, $P(n)$, by engineering filter functions. 
    (a) A family of filter functions, $F_j(n)$, is designed to probe distinct regions in Fock space. 
     Each filter yields a measurement outcome $M_j$.
     $P(n)$ is recovered from the set of filters and corresponding measurements.
    (b) Filters are engineered from laser-induced spin-motion interactions. A laser detuned by $k\omega$ from the spin's transition frequency drives the $k$th sideband transition (see inset) with Rabi frequency $\Omega_k(n)$ (plotted from Eq.~\ref{eq:omega_k} with $\eta = 0.085$). Out of the LD regime (beyond the shaded region), the couplings become highly nonlinear and can develop near degeneracies.
    (c--e) Composite-pulse sequences produce narrowband filters that map into Fock space, whose shape depends strongly on the sideband order. The filter responses can be tuned by varying the pulse duration $T$. Combining filters from multiple sideband orders can yield a well-conditioned measurement family for reconstructing $P(n)$.
    }
    \label{fig:figure_1}
\end{figure*}

Reconstructing the phonon-number distribution, $P(n) = \bra{n}\hat{\rho}\ket{n}$, corresponds to estimating the diagonal elements of the motional density matrix, $\hat{\rho}$, in the Fock basis. Within a filter-function framework, each measurement is given by $M_j = \sum_{n=0}^{\infty} F_j(n) P(n)$, where $F_j(n)$ is a filter function which describes the response of measurement $j$ to phonon number $n$. Determining $P(n)$
thus reduces to estimating $\mathbf{P}$ from
\begin{equation}
    \mathbf{F} \mathbf{P} = \mathbf{M},
    \label{eq:p=fm}
\end{equation}
where $\mathbf{P} = \{P(n)\}$ is the vector of phonon populations, $\mathbf{F} = \{F_j(n)\}$ is a filter matrix, and $\mathbf{M} = \{M_j\}$ is the vector of experimentally measured probabilities. Formulating phonon-number reconstruction this way lets us leverage the extensive literature on filter-function design, drawing on its general principles and methodologies developed for a broad range of applications~\cite{alvarez2011, Biercuk2011, Soare2014, Frey2017,Milne2020, Milne2021b, Kang2023b}. 

For reliable reconstruction, the filter matrix must satisfy two conditions. First, $\mathbf{F}$ must have full column rank to provide sufficient independent constraints to distinguish every phonon number in a range. Second, $\mathbf{F}$ should be well-conditioned (i.e., $\mathbf{F}$ has a low condition number), so that small measurement errors in $\mathbf{M}$ introduce correspondingly small errors in estimating $\mathbf{P}$~\cite{Golub2013}.

The filter-function matrix $\mathbf{F}$ can be constructed from experimentally accessible interactions. Here, we use laser-driven spin-motion interactions to map motional information onto the ion's internal spin, which is subsequently measured by state-dependent fluorescence. As illustrated in Fig.~\ref{fig:figure_1}(a), each experimental setting $j$ applies a spin-motion unitary, $\hat{U}_j$, followed by a spin measurement. This sequence defines the filter function,
\begin{align}
   & F_j(n) = \sum_{m=0}^{\infty} \big|\bra{\uparrow, m} \hat{U}_j \ket{\downarrow, n} \big|^2, 
   \label{eq:filter_function}
\end{align}
which gives the probability of measuring the spin in $\ket{\uparrow}$ for an initial state, $\ket{\downarrow, n}$. 

The spin-motion unitaries, $\hat{U}_j$, are implemented with a laser field detuned by $k \omega$ from the spin's transition frequency, where $\omega$ is the motional frequency and $k \in \mathbb{Z}$ labels the sideband order (e.g.\ $k=0$ for carrier, $k=1$ for first blue-sideband, $k=-2$ for second red-sideband). In the following, we restrict to $k \geq 0$ without loss of generality. The interaction-picture Hamiltonian after making the rotating-wave approximation is
\begin{equation}
    \hat{H}_k(t) = \frac{i^k}{2} e^{- i \phi(t)} \sum_{n=0}^{\infty} \Omega_k(n) \ket{\uparrow, n+k}\bra{\downarrow, n} + \mathrm{h.c.},
    \label{eq:H_k(t)}
\end{equation}
where $\phi(t)$ is a controllable laser phase. Applying $\hat{H}_k(t)$ drives transitions between states $\ket{\downarrow, n} \rightarrow \ket{\uparrow, n+k}$ with coupling strength~\cite{Wineland1998}
\begin{align}
    \Omega_k(n) = \Omega_0 \eta^k e^{-\eta^2/2} \sqrt{\frac{n!}{(n+k)!}} L_n^{(k)}(\eta^2),
    \label{eq:omega_k}
\end{align}
where $\Omega_0$ is the bare carrier Rabi frequency and $L_n^{(k)}$ is the generalized Laguerre polynomial. The LD parameter $\eta = \kappa z_0$ is determined by the effective laser wavevector, $\kappa$, and the zero-point motional extent, $z_0 = \sqrt{\hbar / (2 m \omega)}$, where $m$ is the ion mass. 

In the LD regime, wherein $\eta^2 (2n + 1) \ll 1$, $\Omega_k(n)$ vary monotonically with $n$ and thus provide distinguishable phonon number dependencies. For example, the $k=\{0,1,2\}$ couplings of Eq.~\ref{eq:omega_k} scale approximately as $\Omega_{k=0}(n) \propto 1$, $\Omega_{k=1}(n) \propto \sqrt{n+1}$ and $\Omega_{k=2} \propto n$, respectively. However, beyond the LD regime at large $n$, these coupling strengths become highly nonlinear and can develop near degeneracies, such that different phonon numbers produce nearly identical measurement responses.
For instance, near $n=115$, the $k=1$ coupling is nearly degenerate over several neighboring phonon numbers (Fig.~\ref{fig:figure_1}(b)), making these states difficult to distinguish.

Existing reconstruction protocols can be understood within the filter-function framework, and generally fail to provide both distinguishability and good conditioning over arbitrary phonon-number ranges. Protocols based on measurements from a single sideband cannot guarantee a full-rank filter matrix, because degeneracies in the nonlinear spin-motion coupling can cause distinct phonon numbers to produce indistinguishable measurements. Previous approaches have extended reconstruction beyond the LD regime using multiple or higher-order sidebands, but they either rely on prior knowledge of the motional state~\cite{Walther2012, Alonso2016} or do not guarantee full rank over arbitrary phonon ranges~\cite{Simoni2025}. Separately, composite-pulse sequences have been used to engineer narrowband responses and improve reconstruction conditioning, but they do not resolve degeneracies in the underlying sideband couplings~\cite{Mallweger2024}.

In this work, we generalize phonon reconstruction to arbitrary phonon ranges by designing filter matrices that are both full rank and well-conditioned. We combine measurements from different sideband orders, $k$, to ensure full rank, and engineer the time-dependent phase, $\phi(t)$, to improve the conditioning by implementing narrowband filters in Rabi-frequency space. Specifically, we implement the composite-pulse (CP) sequence of Ref.~\cite{Torosov2015, Mallweger2024} of total duration $T$. It consists of $J$ phase segments of equal duration, whose phases are chosen to produce a narrow passband centred at $\Omega_k T/J ~ \mathrm{mod}~2\pi = \pi$ (see Supplemental Material (SM) for details). Increasing $J$ narrows the filter bandwidth at the cost of longer total duration $T$, and we empirically find that $J=7$ provides a suitable balance.

\begin{figure*}[t]
    \centering
    \includegraphics[]{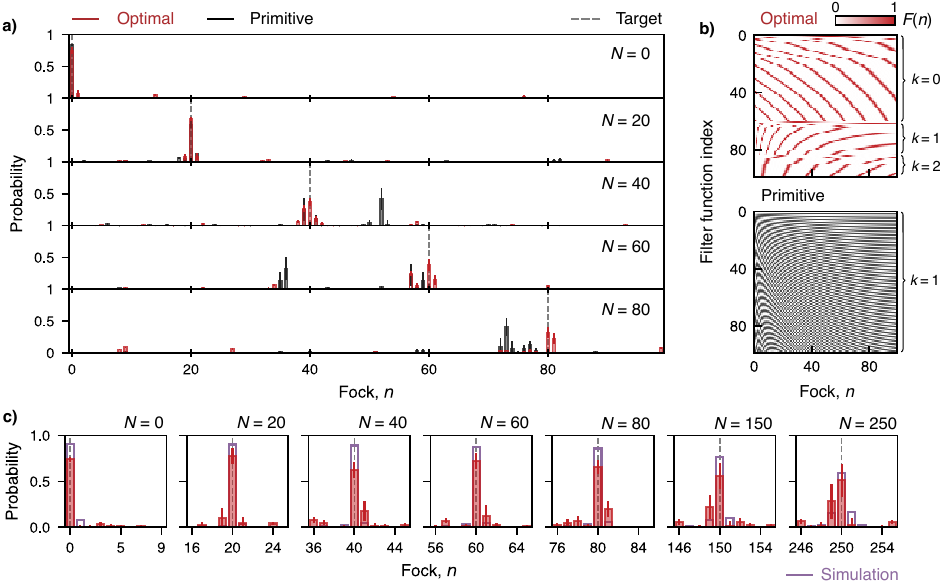}
    \caption{ Experimental reconstruction of Fock states.
    (a) Reconstructed phonon distributions, $P(n)$, over the range $n \in [0, 100)$ for target Fock states $N \in \{0, 20, 40, 60, 80\}$ (dashed lines). The same filter matrix is used across all target Fock states: an optimal filter family (red) is compared to the primitive first blue-sideband reconstruction without composite-pulse modulation (gray). 
    (b) Filter matrices used for the reconstruction of Fock states in (a). (Top) The optimal protocol combines $k=\{0,1,2\}$ filters, whereas the primitive protocol (bottom) uses only $k=1$ filters. The sparsity in the optimal filter matrix and the distinguishability of columns indicate a better-conditioned matrix than the primitive case. 
    (c) Narrow-range reconstructions using filter families separately optimized for phonon ranges around each target Fock state. The narrower reconstruction range reduces reconstruction errors, improving agreement with the expected distribution (purple) obtained from numerical simulations that include noise. Numerical simulations model motional decoherence during preparation pulses, and consider an initial thermal state with $\mathbb{E}[n] = 0.1$ and Lindbladian operators $\sqrt{\gamma_\mathrm{d}}\hat{a}^\dagger \hat{a}$, $\sqrt{\gamma_\mathrm{h}} \hat{a}^\dagger$ and $\sqrt{\gamma_\mathrm{h}} \hat{a}$, with $\gamma_\mathrm{h} = \SI{0.2}{s^{-1}}$ and $\gamma_\mathrm{d} = \SI{18}{s^{-1}}$. Error bars in (a) and (c) denote one standard-deviation of the reconstruction uncertainty, estimated by Monte Carlo propagation of quantum projection noise (QPN) from 150 and 200 measurement repetitions respectively, LD parameter uncertainty and Rabi frequency fluctuations, and this sampling procedure is repeated 100 times (see SM).}
    \label{fig:figure_2}
\end{figure*}

Figure~\ref{fig:figure_1}(c--e) shows how narrowband filters generated from the same CP sequence map into Fock space for different sideband orders. The resulting filter shape is determined by $\Omega_k(n)$ and the total pulse duration, $T$. Because $\Omega_k(n)$ has a distinct phonon-number dependence for each $k$, the corresponding filters probe Fock space differently and can therefore provide complementary selectivity across the target range. For carrier $(k=0)$, increasing $T$ narrows the filter and shifts its response towards larger $n$ (Fig.~\ref{fig:figure_1}(c)). At longer durations, however, degeneracies in $\Omega_0(n)$, together with the periodic filter response, generate additional passbands that cause distinct phonon occupations to produce similar responses within the same filter. For the first-order blue-sideband ($k=1$), the filter response broadens near the first turning point around $n = 115$, where neighboring occupations have nearly identical coupling strengths and are therefore difficult to distinguish (Fig.~\ref{fig:figure_1}(d)). The second-order blue-sideband ($k=2$) produces similar behavior as the first-order sideband, but the turning point is at a higher $n$ (Fig.~\ref{fig:figure_1}(e)).

These examples illustrate the complementary roles of CP duration, $T$, and the sideband order, $k$.  Varying $T$ tunes the location and width of the filter, while $k$ determines how the response maps into Fock space. Combining measurements across different sideband orders therefore provides distinguishability and enables the construction of a full-rank, well-conditioned family of filters tailored to an arbitrary target phonon range.

We demonstrate the reconstruction protocol with a $^{171}$Yb$^+$ ion in a room-temperature Paul trap. We define an effective spin-$\frac{1}{2}$ system in the electronic ground state, with a $T_2^*$ coherence time of $8.9$~s~\cite{Tan2023}. We use a radial motional mode with a frequency $\omega/2\pi = \SI{1.40}{MHz}$, and independently measure the heating and dephasing rates to be $\gamma_\mathrm{h} = \SI{0.2}{s^{-1}}$ and $\gamma_{\mathrm{d}} = \SI{18}{s^{-1}}$, respectively. The Hamiltonian of Eq.~\ref{eq:H_k(t)} is enacted through stimulated Raman transitions using a 355~nm laser; experiments are detailed in Ref.~\cite{Valahu2023, Matsos2024}.  

We optimize a filter family for each target phonon range, $n_\mathrm{min} \leq n  <  n_\mathrm{max}$, to have full column rank and for the total number of filters in each family to be equal to the number of unknown phonon populations, $n_\mathrm{max} - n_\mathrm{min}$. This ensures measurements are informationally complete and uniquely determine the phonon distribution using the minimum number of measurements. Here, filters are selected iteratively using a greedy A-optimal design algorithm~\cite{Pukelsheim1993}, which minimizes the trace of the estimator covariance matrix (i.e.~the average estimation uncertainty) over the target reconstruction range. We restrict the selection to sideband orders $k\in \{0, 1, 2 \}$, as higher-order interactions become prohibitively weak relative to decoherence rates. We also constrain the total duration to $T \leq \SI{3}{ms}$.

\begin{figure}[t]
    \centering
    \includegraphics[width = \columnwidth]{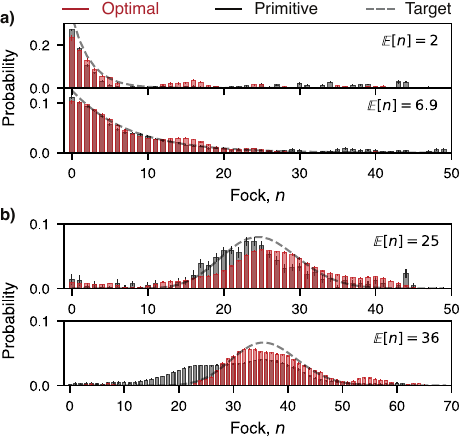}
    \caption{Reconstruction of phonon distributions $P(n)$ for different target mixed and coherent states (dashed line) using composite pulses (red) and primitive pulses (gray). (a) Reconstructed thermal states with target mean phonon numbers $\mathbb{E}[n]= 2 $ and $\mathbb{E}[n] = 6.9$ over the range $n \in [0, 49]$. (b) Reconstructed coherent states with target mean phonon numbers $\mathbb{E}[n] = 25$ ($|\alpha| = 5$) and $\mathbb{E}[n] = 36$ ($|\alpha| = 6$) over the ranges $n \in [0, 49]$ and $n \in [0, 69]$, respectively. Error bars are obtained using the same Monte Carlo uncertainty propagation as in Fig.~\ref{fig:figure_2} with 800 measurement repetitions.}
    \label{fig:figure_3}
\end{figure}

We reconstruct $P(n)$ via constrained convex optimization,
\begin{equation}
    \mathrm{min}_{\mathbf{P}} (||\mathbf{F} \mathbf{P} - \mathbf{M}||^2_2), 
    \label{eq:convexOpt}
\end{equation}
with constraints $\sum_nP(n) \leq 1$ and $0 \leq P(n) \leq 1$; this avoids unphysical solutions that can arise from direct inversion of $\mathbf{F}$ in the presence of measurement noise. 

Our reconstruction protocol relies on accurately implementing the filters, which requires precise calibration of $\eta$ and stabilization of $\Omega_k$. 
We determine $\eta$ by measuring the carrier Rabi frequency, $\Omega_0(n)$, across a range of Fock states; a fit to Eq.~\ref{eq:omega_k} yields $\eta = 0.085$ (see SM). 
We calibrate $\Omega_k$ for each sideband order, $k$, giving nominal Rabi frequencies $\Omega_0(0) / 2\pi = \SI{33}{kHz}$, $\Omega_1(0) /2 \pi = \SI{2.0}{kHz}$ and $\Omega_2(0)/2\pi = \SI{0.25}{kHz}$. We stabilize fluctuations in $\Omega_k$ due to laser intensity noise using a closed-loop servo that exploits the narrowband response of the CP sequence to enhance sensitivity to amplitude drifts (see SM). 

In the first experiment, we investigate the protocol's efficacy by reconstructing Fock states in the range $0 \leq n < 100$ (see Fig.~\ref{fig:figure_2}(a)). Target Fock states are prepared by first cooling to the motional ground state, followed by alternating blue- and red-sideband $\pi$-pulses that sequentially increase $n$. Mid-circuit measurements are interleaved throughout the preparation sequence to suppress errors arising from imperfections in laser pulses (see SM). The optimized filter family for this range comprises 60 $k=0$, 24 $k=1$, and 16 $k=2$ filters (Fig.~\ref{fig:figure_2}(b)). 

\begin{figure}[t]
    \centering
    \includegraphics[width=\columnwidth]{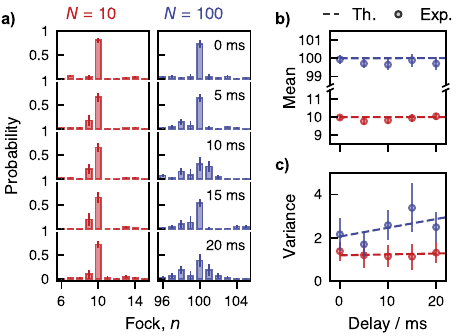}
    \caption{Heating dynamics of Fock states. (a) Reconstructed phonon distributions, $P(n)$, for target Fock states $N=10$ (left, red) and $N=100$ (right, blue) after variable delays following state preparation. Reconstructions use filter-function families consisting of $k=0$ filters optimized for their corresponding ranges $n\in[6, 15]$ and $n\in[96, 105]$, respectively. Error bars are obtained using the same Monte Carlo uncertainty propagation as in Fig.~\ref{fig:figure_2} with 100 measurements for QPN. (b) Mean phonon number, $\mathbb{E}[n]$, and (c) variance, $\mathbb{V}[n]$, extracted from the reconstructed distributions. Dashed lines show theoretical predictions using an independently calibrated heating rate.}
    \label{fig:figure_4}
\end{figure}

Figure~\ref{fig:figure_2}(a) shows that the reconstruction protocol correctly identifies the target Fock state across the full range. The reconstruction quality gradually decreases for higher target Fock states, consistent with the intrinsic estimation uncertainty of the filter matrix, $\sqrt{\mathrm{diag}\left((\mathbf{F}^{\mathsf{T}} \mathbf{F})^{-1}\right)}$, rather than imperfections in the state preparation (see SM). We find that this filter-function matrix gives uncertainties that grow with $n$.

To verify that this degradation originates from the filter design, we repeat the reconstructions over narrower phonon ranges, and optimize a new filter family comprising 10 $k=0$ filters for each range (Fig.~\ref{fig:figure_2}(c)). The resulting filter families exhibit substantially lower estimation uncertainty (see SM), yielding reconstructions in significantly better agreement with the ideal target states. We further perform a global sensitivity analysis and find that, for the $N=250$ state, reconstruction uncertainty is dominated by QPN and uncertainty in the calibrated LD parameter (see SM).

We compare our protocol against a `primitive' reconstruction protocol that uses the $k=1$ transition without a CP sequence (Fig.~\ref{fig:figure_2}(a), gray), which is commonly used to reconstruct $P(n)$ within the LD regime~\cite{Meekhof1996, Leibfried1996}. We construct the primitive filter matrix $\mathbf{F}_\mathrm{prim}$ by uniformly sampling the pulse durations over the maximum duration used by the optimized filter family. The reconstructed $P(n)$ agree with the ideal target state $N=20$, beyond which they significantly deviate. This breakdown directly reflects the properties of $\mathbf{F}_\mathrm{prim}$ (Fig.~\ref{fig:figure_2}(b)): first, $\mathbf{F}_\mathrm{prim}$ is rank-deficient, indicating that the measurements are informationally incomplete; second, the estimation uncertainty significantly increases for $n\geq 25$, indicating ill-conditioning beyond this range. 

Next, we reconstruct $P(n)$ of thermal and coherent states (Fig.~\ref{fig:figure_3}). Each state is prepared by applying an oscillating electric field resonant with $\omega$ (see SM). Filter functions are optimized using 50 and 70 $k=0$ filters over the range $0 \leq n <50$ and $0 \leq n <70$, respectively. For these reconstructions, we further penalize unphysical oscillatory solutions arising from amplification of measurement noise as they become increasingly dominant for these wide-range filter families \cite{Hansen1994}. This is done by adding a regularization term in Eq.~\ref{eq:convexOpt} (see SM). 

The optimal protocol yields reconstructed distributions that show good agreement with all ideal target states. In contrast, reconstructions using the primitive protocol degrade beyond the LD regime, as apparent for the coherent state with $\mathbb{E}[n]=36$ (gray, Fig.~\ref{fig:figure_3}b). This behavior is consistent with the filter matrix being rank-deficient, which makes measurements informationally incomplete. The secondary lobes that appear in the optimal reconstructions can be attributed to the non-negligible off-diagonal elements of the associated cosine-similarity matrix (see SM). In general, reconstruction performance can be systematically improved by engineering narrower filters through CP sequences with more segments, longer pulse durations, and incorporating additional high-order sideband interactions. 

Our protocol also enables measurements of highly excited motional dynamics. We measure the motional heating of Fock states $N=10$ and $N=100$ (Fig.~\ref{fig:figure_4}). A varying delay of up to $\SI{20}{ms}$ is added between state preparation and reconstruction. Figure~\ref{fig:figure_4}(a) shows that the phonon distributions broaden with increasing delay, with a rate that increases with $N$.
We plot the evolution of the mean and variance of the distributions in Fig.~\ref{fig:figure_4}(b,c) and compare with a model that includes bosonic loss and gain channels described by $\hat{L}_1 = \sqrt{\gamma_\mathrm{h}} \hat{a}$ and $\hat{L}_2 = \sqrt{\gamma_\mathrm{h}} \hat{a}^\dagger$. Under this model, the mean phonon number of an initial Fock state $\ket{N}$ increases linearly at a rate independent of $N$, $d\mathbb{E}[n]/dt = \gamma_\mathrm{h}$, while the variance grows at a rate $d\mathbb{V}[n]/dt = \gamma_\mathrm{h} (2N + 1)$. Our measurement (circles) shows good agreement with the model (dashed lines).

In summary, we introduce and experimentally demonstrate an optimal filter-function framework for reconstructing the probability distributions of trapped-ion motional states beyond the LD regime. By combining composite-pulse control across multiple orders of spin-motion interactions, the method constructs full-rank, well-conditioned measurement families. Our protocol is applicable to both characterization and dynamical studies of arbitrary motional states over arbitrary phonon-number ranges. We demonstrate its performance on multiple pure and mixed states, with higher reconstruction fidelities than a primitive protocol of the same duration. This work also presents the experimental realization and verification of Fock states as high as $N = 250$. 

The reconstruction protocol can serve as a thermometry technique without assuming the expected phonon distribution a priori and is applicable beyond the LD regime, making it useful for systematic evaluations of precision spectroscopy and atomic clocks~\cite{Chen2017, King2022}. 

Looking forward, the filter-function framework can be extended to a broader class of spin-motion unitaries to enrich filter designs, and enable reconstructions of full density matrices beyond the LD regime. These capabilities would open a route towards using highly excited motional quantum states in quantum sensing, simulation, information processing, and for studying quantum mechanics at high excitations.

We thank Liam Bond, Matteo Simoni, Wojciech Adamczyk, Moritz Fontbot\'e-Schmidt, Jeremy Metzner, and Jonathan Home for useful discussions. We were supported by the U.S. Office of Naval Research Global (N62909-24-1-2083), the U.S. Air Force Office of Scientific Research (FA2386-23-1-4062), the Australian Research Council (FT220100359, DP260104144), the Sydney Quantum Academy (MJM, PN), the University of Sydney Postgraduate Award scholarship (VGM, HKC), the Australian Government Research Training Program (FS), the Sydney Horizon Fellowship (TRT), and H.\ and A.\ Harley.

The data that support the findings of this study are publicly available~\cite{Valahu2026}.

\section{Supplemental Material}



\subsection{Lamb-Dicke parameter calibration}

\setcounter{equation}{0}
\renewcommand{\theequation}{SM\arabic{equation}}
\setcounter{figure}{0}
\renewcommand{\thefigure}{SM\arabic{figure}}

Reliable phonon-number reconstruction beyond the Lamb-Dicke (LD) regime requires an accurate calibration of the LD parameter, since the nonlinear dependence of the sideband couplings on $\eta$ becomes increasingly pronounced at large phonon numbers. We determine $\eta$ by fitting the carrier Rabi frequency to Eq.~4 of the main text with $\Omega_0(n)$ measured over a range of prepared Fock states. To maximize sensitivity, we target Fock states near the first zero-crossing of $L_n^{(0)}(\eta^2)$. We repeat this calibration daily to compensate for slow drifts of the motional frequency.

\begin{figure}[t]
    \centering
    \includegraphics[]{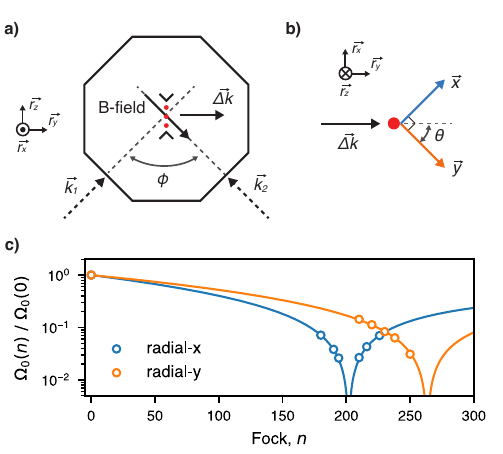}
    \caption{Calibration of the Lamb-Dicke parameters for radial modes $x$ and $y$. (a) The Lamb-Dicke parameter is proportional to the projection of the wavevector with the normal mode. Light-atom interactions are enacted by stimulated Raman transitions from two Raman beams with wavevectors $\vec{k}_1$ and $\vec{k}_2$, giving a difference wavevector $\vec{\Delta k} = \vec{k}_1 - \vec{k}_2$. The angle between the two Raman beams $\phi$ is ideally $90^\circ$. (b) In the radial plane, the two radial modes along directions $\vec{x}$ and $\vec{y}$ are orthogonal. The difference wavevector is at an angle $\theta$ with respect to $\vec{y}$, which should ideally be $45^\circ$. (c) The Lamb-Dicke parameters for each mode are obtained by measuring the carrier Rabi frequency, $\Omega_0(n)$, for different initial Fock states, $n$ (markers). The Lamb-Dicke parameters are extracted from fits to theory (solid lines, cf. Eq.~4).}
    \label{fig:ld_fitting}
\end{figure}

Figure~\ref{fig:ld_fitting}(c) shows measurements of carrier Rabi frequencies for both the radial-$x$ and -$y$ motional modes. Fits to Eq.~4 yield $\eta_x = 0.08453(3)$ and $\eta_y = 0.07401(5)$, with corresponding motional frequencies $\omega_x/2\pi = \SI{1.400}{MHz}$ and $\omega_y/2\pi = \SI{1.527}{MHz}$, respectively. 

We compare the measured LD parameters to theoretical predictions derived from the Raman laser beam geometry (Fig.~\ref{fig:ld_fitting}(a,b)). The LD parameter of the $i$th mode is 
\begin{equation}
    \eta_i = k_i z_{0,i},
\end{equation}
where $k_i = \vec{\Delta k} \cdot \vec{i}$ is the projection of $\vec{\Delta k}$ onto the $i$th mode, with $i\in \{ x,y,z\}$. The Raman wavevector difference is $\vec{\Delta k} = \vec{k}_1 - \vec{k}_2 = \frac{4 \pi}{\lambda} \sin(\frac{\phi}{2})$ while $z_{0,i} = \sqrt{\hbar/(2 m \omega_i)}$ is the ground state spatial extent. As illustrated in Fig.~\ref{fig:ld_fitting}(a,b), we parameterize the geometry by two angles: $\phi$, the angle between the two Raman beams (ideally $90^\circ$), and $\theta$, the angle between $\vec{\Delta k}$ and mode $y$ (ideally $45^\circ$). With this, the LD parameters can be expressed as
\begin{align}
    & \eta_x = \sin\left(\frac{\phi}{2}\right)\sin(\theta) \sqrt{\frac{2 \hbar}{m \omega_x}} \frac{2\pi}{\lambda}, \label{eq:eta_x} \\
    & \eta_y =  \sin\left(\frac{\phi}{2}\right)\cos(\theta) \sqrt{\frac{2 \hbar}{m \omega_y}}  \frac{2\pi}{\lambda}. \label{eq:eta_y}
\end{align}

Substituting the fitted LD parameters into Eq.~\ref{eq:eta_x} and Eq.~\ref{eq:eta_y} gives $\phi = 89.54(4)^\circ$ and $\theta = 47.57(2)^\circ$, which is in good agreement with the ideal geometry. 
\subsection{Composite-pulse sequence}

\begin{figure}
    \centering
    \includegraphics[]{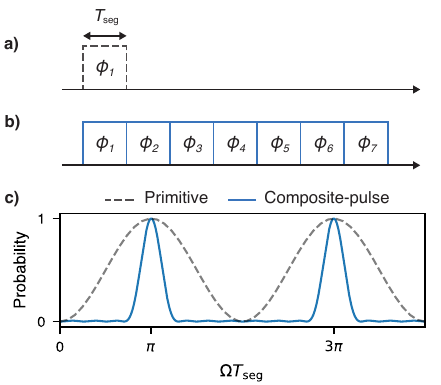}
    \caption{Composite-pulse filter.
    (a) Primitive pulse with constant phase and duration $T_\mathrm{seg}$.
    (b) Composite-pulse consisting of $J=7$ phase segments of equal duration, $T_\mathrm{seg}$.
    (c) Spin-excitation probability as a function of the dimensionless pulse area, $\Omega T_\mathrm{seg}$ for the primitive pulse (dashed) and the composite pulse (solid). The phases of the composite pulse are those used in the experiments of the main text.}
    \label{fig:composite_pulse}
\end{figure}

We engineer narrowband filter functions by phase modulating the spin-motion interaction using the composite-pulse (CP) sequence introduced in Refs.~\cite{Torosov2015, Mallweger2024}. This CP sequence implements a piecewise constant phase modulation consisting of $J$ segments of equal duration, $T_\mathrm{seg}$, such that the total interaction time is $T=J T_\mathrm{seg}$. The phases that make up each segment of the sequence are optimized such that the resulting spin-excitation probability forms a narrow passband filter as a function of Rabi frequency for a fixed $T$. The passband centres occur at
\begin{equation}
    \Omega_k T_\mathrm{seg} = \pi (1 + 2m), ~ ~ ~ m=0, 1, 2, ...
    \label{eq:pulse_area}
\end{equation}
For a sideband interaction of order $k$, the Rabi frequency depends on the initial Fock state through $\Omega_k(n)$ of Eq.~4 of the main text. Consequently, the narrowband response in Rabi-frequency space is mapped onto a filter in Fock space, whose shape is determined by $\Omega_k(n)$.

Figure~\ref{fig:composite_pulse} plots the spin excitation response of the CP sequence. A primitive pulse with $J=1$ implements an effective Rabi oscillation, producing a sinusoidal excitation. A CP sequence with $J=7$ narrows the excitation profile around each passband centre. Further increasing $J$ reduces the bandwidth of the CP filter, at the cost of increasing the total interaction duration, $T$.

For all CP sequences used in the main text, we choose $J=7$ with phases $\pi \times \{0, 0.299, 0.972, 0.850, 0.727, 1.400, 1.700\}$, taken from Ref.~\cite{Mallweger2024}.

\subsection{Rabi frequency stabilization}

Accurate reconstruction requires that the Rabi frequency remain stable throughout the protocol. Since the passband centres of the CP sequence depend on the pulse area, $\Omega_k T_\mathrm{seg}$, fluctuations in $\Omega_k$ shift the filter response and lead to errors in the implemented filter functions. We therefore frequently calibrate the Rabi frequency using a closed-loop servo algorithm originally devised for stabilizing laser frequencies~\cite{Oskay2005, Peik2005}. The servo exploits the narrow passbands of the CP sequence to provide enhanced sensitivity to amplitude drifts.

The stabilization algorithm implements a two-point balanced servo on the spin-excitation of the CP sequence. We probe the response on either side of a selected passband and use the difference in spin-excitation to construct an error signal. This error signal provides an estimate of the Rabi-frequency drift, and is used to update the pulse durations, such that the total pulse area remains constant.

Specifically, to stabilize the Rabi frequency $\Omega_k$ of the $k$th interaction order, we initialize the spin and motion to their ground state, $\ket{\downarrow, 0}$, and apply the CP sequence with $J=7$ phase segments. We measure the $m$th passband by setting the segment duration to $T_\mathrm{seg} = \pi (1 + 2m)/ \Omega_k$ (Eq.~\ref{eq:pulse_area}). To generate an error signal, we probe the two sides of the passband at segment durations $T_\mathrm{seg} \pm \delta T$, yielding spin-excitation probabilities $p_\pm$. We determine $\delta T$ experimentally such that the two probe points lie at the half-maximum of the passband. In the absence of Rabi frequency fluctuations, the probe points are symmetric about the passband centre and therefore give $p_+ = p_-$. However, changes in $\Omega_k$ shift the passband relative to the probe points and produce an imbalance in the probabilities.

We construct a normalized error signal from the measured probabilities,
\begin{equation}
    \epsilon = g\frac{p_+ - p_-}{p_+ + p_-},
\end{equation}
where $g$ is the servo gain. Normalizing by $(p_+ + p_-)$ reduces sensitivity to variations in the overall spin excitation contrast. After each servo measurement, the estimated Rabi frequency is updated according to $\Omega_k \rightarrow \Omega_k (1 - \epsilon)$. The segment duration of the CP sequence is updated accordingly, such that the pulse area of a target filter function remains unchanged.

In our experiment, we operate the servo on the $m=10$ passband for the carrier ($k=0$), and on the $m=5$ passband for the first- ($k=1$) and second-order ($k=2$) sidebands. Additionally, prior to applying the servo, the motional state is initialised in the ground state for the carrier and first-order sidebands, and on a Fock $N=10$ state for the second-order sideband.
The gain is set to $g = g_0 / (1 + 2m)$, where $g_0 = 0.005$ is determined empirically. Each of the probabilities $p_+$ and $p_-$ is estimated from 25 measurement repetitions. 

\begin{figure}[t]
    \centering
    \includegraphics[]{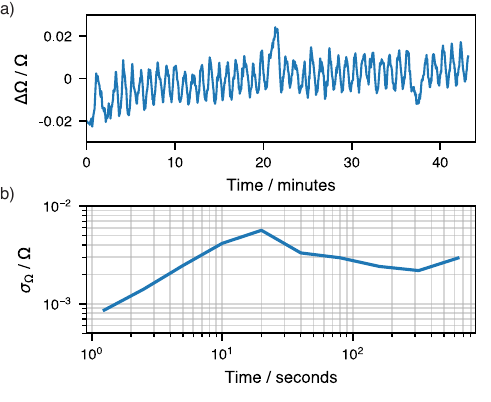}
    \caption{Long-term stability of the Rabi frequency, $\Omega$.(a) Fractional deviation $\Delta\Omega / \Omega$ over a 43 minute period of the carrier Rabi frequency with initial Fock state $N=0$. At each servo step, the probabilities were measured with $25$ measurement repetitions each, and the target peak was $m=10$. (b) Corresponding Allan deviation of the Rabi frequency time series. }
    \label{fig:omega_drift}
\end{figure}

Figure~\ref{fig:omega_drift} characterizes the long-term stability of the carrier Rabi frequency, $\Omega_{k=0}(n=0)$, using the above servo algorithm. The Rabi frequency exhibits a dominant oscillation with a period of 54 seconds and a fractional amplitude of approximately $0.5\%$. The corresponding Allan deviation reaches a minimum fractional instability of $0.09\%$ for an averaging time of $\SI{1.2}{s}$. During phonon-distribution measurements, the servo is applied at an interval that is much shorter than $54$ seconds to compensate for slow drifts in $\Omega_k$.

\subsection{Fock state preparation}

\begin{figure}[t]
    \centering
    \includegraphics[]{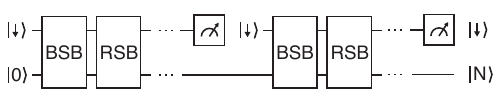}
    \caption{Pulse sequence for preparing a target Fock state $\ket{n}$. After initializing in the $\ket{\downarrow}\ket{0}$ state, an alternating sequence of blue- and red-sideband pi pulses sequentially increases the Fock number. Interleaved mid-circuit spin measurements ensure state purity.}
    \label{fig:fock_prep}
\end{figure}

We prepare Fock states by first initializing the system to $\ket{\downarrow}\ket{0}$, followed by applying an alternating sequence of blue-sideband ($k=1$) and red-sideband ($k=-1$) pulses, which sequentially increase the Fock number by 1 (Fig.~\ref{fig:fock_prep}). Blue-sideband pulses implement the transition $\ket{\downarrow, n} \rightarrow \ket{\uparrow, n+1}$, while red-sideband pulses implement the transition $\ket{\uparrow, n} \rightarrow \ket{\downarrow, n+1}$. The duration of each sideband pulse is set to implement a $\pi$ pulse for each $n$, with duration $T_\pi = \pi / \Omega_k(n)$.

We incorporate mid-circuit spin measurements throughout the sideband pulse sequence, as well as after the final pulse, to suppress Fock-state preparation errors arising from imperfect laser pulses. The sequence proceeds only if we measure $\ket{\downarrow}$, because detection of $\ket{\uparrow}$ indicates unwanted spin-motion entanglement and involves photon scattering that decoheres the motional state due to random momentum kicks. A $\ket{\downarrow}$ measurement does not decohere the motional state because, ideally, no photons are scattered during the measurement.
Incorporating these mid-circuit measurements as the Fock state is progressively prepared along the phonon-number ladder preserves the purity of the motional state at successive stages and improves the overall fidelity of the target Fock state. In practice, we measure the spin state after 10 to 20 pairs of blue- and red-sideband pulses. The total duration for preparing the Fock $N=250$ state is approximately $\SI{8}{ms}$.

\subsection{Preparing coherent and thermal states}

Coherent states are prepared by applying an oscillating voltage to one of the ion trap's compensation electrodes, which produces an oscillating electric field at the ion location. Setting the oscillation frequency resonant with the motional frequency implements the interaction Hamiltonian $H =i \chi (\hat{a}^\dagger e^{i \phi} - \hat{a} e^{-i \phi})/2$, where $\chi/2\pi = \SI{6.2}{kHz}$ is the calibrated coupling strength. Applying this interaction for a duration $t$ creates a coherent state $\ket{\alpha}$ with $ \alpha = \chi te^{i \phi}/2$. 

Thermal states are prepared by creating a Gaussian mixture of coherent states with mean phonon number, $\mathbb{E}[n]$~\cite{Glauber1963}. At each of the 800 shots of the experiment, the phase is uniformly sampled in the range $[0, 2\pi)$, and the displacement magnitude is sampled from a Rayleigh distribution with scale parameter $\sigma = \sqrt{\mathbb{E}[n]/2}$.

\subsection{Global sensitivity analysis}

We perform a variance-based global sensitivity analysis to identify the dominant sources of uncertainty in the narrow-range Fock state reconstructions shown in Fig.~2(c). We consider three sources of uncertainty: fluctuations in the Rabi frequency, uncertainty in the calibrated LD parameter, and quantum projection noise (QPN). For each target Fock state, we independently sample these uncertainties, repeat the reconstruction, and quantify each source's contribution to the variance of the reconstructed phonon distribution.

The uncertainty in the Rabi frequency is dominated by slow laser-intensity fluctuations between successive applications of the stabilization servo described above. We model these fluctuations by sampling the Rabi frequency from a normal distribution with a fractional standard deviation of $10^{-3}$, determined by the mean change between successive Rabi frequency calibrations. We next sample the LD parameter from a normal distribution with a relative standard deviation of $5 \times 10^{-4}$, obtained from the calibration uncertainty (Fig.~\ref{fig:ld_fitting}(c)). Lastly, we model QPN by resampling each measured spin-excitation probability from a binomial distribution with 150 trials, matching the number of repetitions used to estimate each probability.

\newcommand{\tabpad}{\rule{0pt}{2.6ex}}
\begin{table}[t]
    \centering
    \label{tab:sobol_indices}
    \renewcommand{\arraystretch}{1.1}
    \begin{tabular*}{\columnwidth}{@{\extracolsep{\fill}}lccc}
        \hline
        \tabpad & \multicolumn{3}{c}{Uncertainty source} \\
        \cline{2-4}
        \tabpad Target state & Rabi frequency & Lamb-Dicke & QPN \\
        \hline
        \tabpad $n=0$ & 64\% & 0\% & 38\% \\
        \tabpad $n=20$ & 80\% & 3\% & 30\% \\
        \tabpad $n=40$ & 59\% & 15\% & 29\% \\
        \tabpad $n=60$ & 47\% & 30\% & 36\% \\
        \tabpad $n=80$ & 38\% & 25\% & 46\% \\
        \tabpad $n=150$ & 11\% & 33\% & 64\% \\
        \tabpad $n=250$ & 8\% & 55\% & 63\% \\
        \hline
    \end{tabular*}
    \caption{Global sensitivity analysis for phonon reconstructions. The table reports the total-order Sobol' indices which quantify the contribution of LD parameter uncertainty, Rabi-frequency fluctuations and quantum projection noise (QPN) to the reconstruction variance. The analysis is performed on the reconstructed Fock states of Fig.~2(c).
    }
\end{table}

Table~\ref{tab:sobol_indices} reports the grouped total-order Sobol' indices associated with each uncertainty source. The total-order index measures the fraction of output variance that can be attributed to a given uncertainty source, including both its direct contribution and its interactions with other sources. The indices show that the dominant source of uncertainty changes substantially with the target phonon number. For states at low phonon numbers, the reconstruction is most sensitive to Rabi-frequency fluctuations, whereas uncertainty in the LD parameter contributes negligibly. For states at higher phonon numbers, the contribution from Rabi-frequency fluctuations decreases, while those from QPN and the LD parameter increase.

These trends follow directly from the behavior of the carrier Rabi frequency, $\Omega_0(n)$, and from the conditioning of the filter matrices. At low $n$, the carrier Rabi frequencies of neighboring phonon numbers are closely spaced, so a small fractional error in the Rabi frequency appreciably shifts the corresponding filters in Fock space and produces a large reconstruction error. The relative spacings of $\Omega_0(n)$ increase with $n$, thereby reducing this sensitivity at higher phonon numbers. Conversely, the nonlinear dependence of $\Omega_0(n)$ on the LD parameter becomes increasingly pronounced at high phonon numbers (cf. Eq.~4), making uncertainties in $\eta$ progressively more important. Finally, the increasing contribution from QPN reflects the conditioning of the optimized filter families which worsens at large $n$: measurement noise is increasingly amplified by the reconstruction, consistent with the covariance and condition-number analysis presented below.

\subsection{Analysis of filter function matrices}

\begin{figure}
    \centering
    \includegraphics[width=\columnwidth]{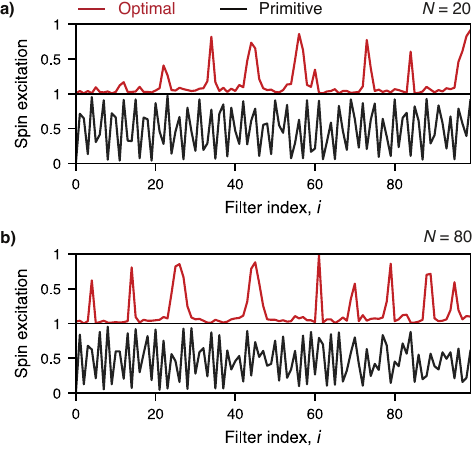}
    \caption{Example of measured spin excitation probabilities as a function of filter function index, $i$, for Fock states (a) $N=20$ and (b) $N=80$, using the optimal (red) and primitive (black) filter functions. These measurements constitute the vector $\mathbf{M}$ of Eq.~1, from which the phonon distributions shown in Fig.~2(a) are reconstructed via the convex optimization procedure of Eq.~5.}
    \label{fig:raw_measurements}
\end{figure}

The quality of the phonon-distribution reconstruction is determined by two properties of the filter matrix, $\mathbf{F}$: its ability to distinguish different Fock states, and its sensitivity to measurement noise. Representative examples of the experimentally measured spin excitation probabilities that form the measurement vector $\mathbf{M}$ are shown in Fig.~\ref{fig:raw_measurements}. Here, we characterize the properties of the filter functions used throughout the main text using several metrics: Fig.~\ref{fig:ff_matrices} shows the filter matrices, their cosine-similarity matrix and their corresponding reconstruction uncertainties, while Table~\ref{tab:condition-uncertainty} summarizes their condition number and mean reconstruction uncertainty.

We first quantify the distinguishability of a filter matrix $\mathbf{F}$ to different Fock states using the cosine-similarity matrix, $\mathbf{S}$, with elements 
\begin{equation}
    S_{ij} = \frac{F^\mathrm{T}_i F_j}{||F_i||~||F_j||},
\end{equation}
where $F_i$ is the vector of filter responses associated with Fock state $i$. Since filter responses are non-negative, the elements of the cosine-similarity matrix lie between $0 \leq S_{ij} \leq 1$. A value of $S_{ij} = 0$ indicates that states $i$ and $j$ produce well-separated measurement responses, whereas $S_{ij} = 1$ indicates identical responses and hence are indistinguishable. Therefore, an ideal filter-function matrix would have $\mathbf{S} = \mathbf{I}$. The off-diagonal structures of $\mathbf{S}$ visible in Fig.~\ref{fig:ff_matrices} reveal pairs or groups of Fock states with similar measurement signatures and, consequently, reduced distinguishability. In the presence of measurement noise, population can therefore be more readily misassigned among these states during reconstruction. This is seen for states in Fig.~3(a,b) where the optimal reconstructions exhibit a secondary lobe due to the non-negligible off-diagonal elements present in their corresponding cosine-similarity matrix, as seen in Fig.~\ref{fig:ff_matrices}(a,b). 

We further quantify the reconstruction uncertainty for each Fock state from the diagonal of the covariance matrix, $\sqrt{\mathrm{diag}\left((\mathbf{F}^\mathrm{T} \mathbf{F})^{-1}\right)}$. Larger values indicate that measurement noise is strongly amplified when estimating $P(n)$. Fig.~\ref{fig:ff_matrices} shows that this uncertainty generally increases with phonon number within a given range (Fig.~\ref{fig:ff_matrices}(a-c)), and that the mean uncertainty is greater for larger phonon ranges (Fig.~\ref{fig:ff_matrices}(d-l)).  

\begin{table}[t]
    \centering
    \begin{tabular}{ccc}
        \hline
        Range & Condition number & Mean uncertainty \\
        \hline
        $[0, 9]$ & $17$ & $2.3$ \\
        $[6, 15]$ & $29$ & $3.1$ \\
        $[16, 25]$ & $29$ & $2.9$ \\
        $[36, 45]$ & $34$ & $3.7$ \\
        $[56, 65]$ & $62$ & $5.6$ \\
        $[76, 85]$ & $1.4 \times 10^{2}$ & $11$ \\
        $[96, 105]$ & $1.7 \times 10^{2}$ & $13$ \\
        $[146, 155]$ & $2.2\times 10^{3}$ & $1.6\times 10^{2}$ \\
        $[246, 255]$ & $8.6 \times 10^{4}$ & $5.8\times 10^{3}$ \\
        $[0, 49]$ & $2.3\times 10^{2}$ & $5.0$ \\
        $[0, 69]$ & $1.2 \times 10^{5}$ & $9.5 \times 10^{2}$ \\
        $[0, 99]$ & $1.8 \times 10^{4}$ & $80$ \\
        \hline
    \end{tabular}
    \caption{Condition number and mean reconstruction uncertainty calculated for each optimal filter function matrix used in the main text. The mean uncertainty is calculated from the mean of the diagonal of the covariance matrix.}
    \label{tab:condition-uncertainty}
\end{table}

In Table \ref{tab:condition-uncertainty}, we report the condition number and the mean uncertainty. The condition number is calculated as $\mathrm{cond}(\mathbf{F}) = \sigma_\mathrm{max}/ \sigma_\mathrm{min}$, where $\sigma_\mathrm{max}$ and $\sigma_\mathrm{min}$ are the largest and smallest singular values of $\mathbf{F}$. Ideally, $\mathrm{cond}(\mathbf{F}) = 1$, and a larger condition number signifies that measurement noise is more amplified during reconstruction. 

These metrics reveal a trade-off between the target phonon range and the reconstruction's quality. For the narrow-range Fock state reconstructions, both the conditioning and reconstruction uncertainty generally worsen as the target phonon number increases. For example, the filter family spanning $n\in [6, 15]$ has $\mathrm{cond}(\mathbf{F}) = 29$ and a mean uncertainty of 3.1, whereas the family spanning $n \in [246, 255]$ has $\mathrm{cond}(\mathbf{F}) = 8.6\times 10^{4}$ and a mean uncertainty $5.8 \times 10^{3}$. This increased noise amplification reflects the growing contribution of quantum projection noise at large $n$, as the global sensitivity analysis above shows. A similar tradeoff arises as the reconstruction range broadens: distinguishing more populations generally worsens the conditioning of the reconstruction. 

\begin{figure}
    \centering
    \includegraphics[width=\columnwidth]{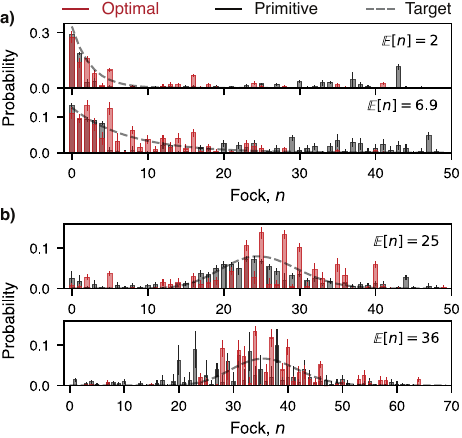}
    \caption{Unregularized reconstructions of phonon distributions for (a) thermal and (b) coherent states using composite pulses (red) and primitive pulses (gray). The regularized reconstructions are shown in Fig.~3.}
    \label{fig:unreg_fig3}
\end{figure}

\begin{figure*}
    \centering
    \includegraphics[width = \textwidth]{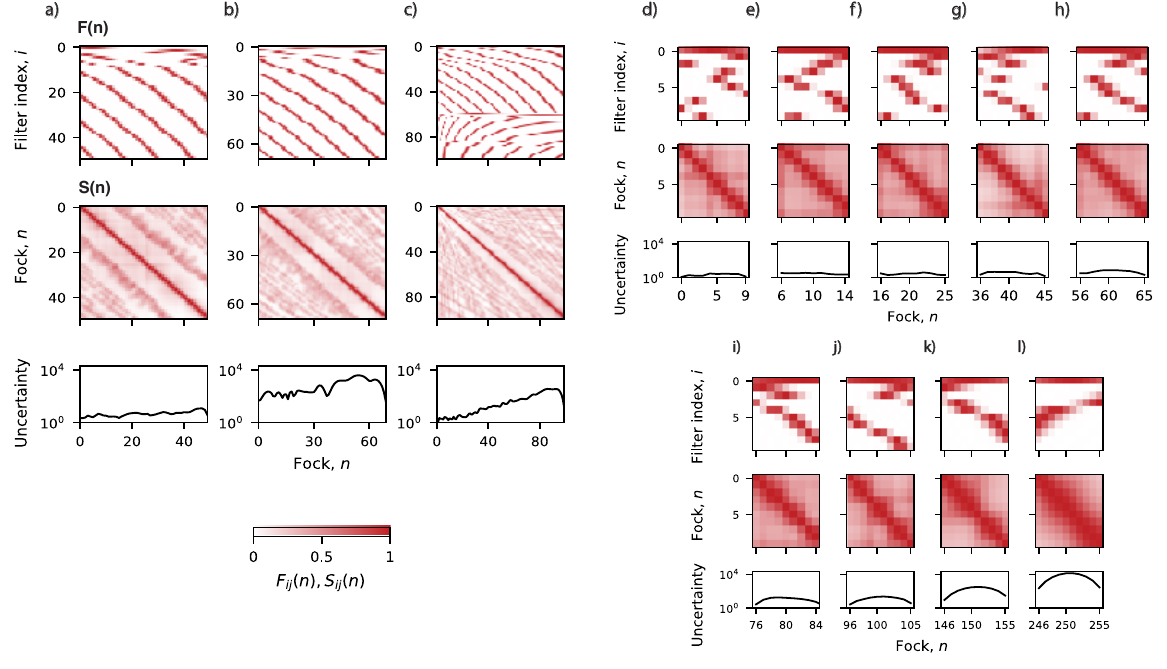}
    \caption{Filter matrices, $\mathbf{F}$, their corresponding cosine similarity matrices, $\mathbf{S}$, and the estimation uncertainty for filter families used in the main text. (a, b) Filters constructed using only carrier ($k = 0$) pulses for relatively wide phonon ranges used for thermal and coherent state reconstructions in Fig.~3. (c) Filters constructed for a wide phonon range $[0,99]$ combining carrier ($k = 0$), first- and second-order ($k = 1, 2$) blue-sideband pulses. They are used for reconstruction of Fock states in Fig.~2(a). (d--l) Narrow-range filters constructed using only carrier ($k = 0$) pulses. In (d), (f--i) and (k--l), filters are used for narrow-range reconstruction of Fock states in Fig.~2(c). Filters in (e) and (j) are used to characterize heating dynamics of Fock states in Fig.~4.}
    \label{fig:ff_matrices}
\end{figure*}

\subsection{Regularized phonon reconstructions of thermal and coherent states}

The thermal and coherent state reconstructions in Fig.~3 employ regularization to suppress large oscillations in $P(n)$ arising from amplification of measurement noise by small singular values of the filter matrix \cite{Tikhonov1963}. This issue is particularly pronounced for reconstructions over broad phonon ranges. The spurious oscillations arise because the right singular vectors associated with small singular values typically exhibit increasingly rapid variations across the phonon range~\cite{Hansen1994}. Indeed, the oscillations observed in the unregularized reconstructions of Fig.~\ref{fig:unreg_fig3}(b) closely align with those of the right singular vectors associated with the smallest singular values of the corresponding filter matrix.

\begin{table}[!t]
    \centering
    \vspace{15pt}
    \renewcommand{\arraystretch}{1.5}
    \begin{tabular*}{\columnwidth}{@{\extracolsep{\fill}}lrrrrrr}
        \hline
        \tabpad & \multicolumn{2}{c}{Target} & \multicolumn{2}{c}{Optimal} & \multicolumn{2}{c}{Primitive} \\
        \cline{2-3} \cline{4-5} \cline{6-7}
        \tabpad State & $\mathbb{E}[n]$ & $\mathbb{V}[n]$ & $\mathbb{E}[n]$ & $\mathbb{V}[n]$ & $\mathbb{E}[n]$ & $\mathbb{V}[n]$ \\
        \hline
        Thermal  & $2$   & $6$     & $6.5(3)$  & $91(8)$  & $10.3(7)$ & $216(14)$ \\
        Thermal  & $6.9$ & $54.51$ & $7.2(3)$  & $45(4)$  & $8.9(6)$  & $110(18)$ \\
        Coherent & $25$  & $25$    & $25.8(4)$ & $78(4)$  & $23.4(6)$ & $74(9)$   \\
        Coherent & $36$  & $36$    & $37.3(4)$ & $75(7)$  & $30.9(7)$ & $114(9)$  \\
        \hline
    \end{tabular*}
    \caption{Characterization of thermal and coherent states reconstructed in Fig.~3 by quantifying their mean phonon number and variance estimated using both optimal and primitive reconstruction methods.}
    \label{tab:fig_3_data}
\end{table}
To suppress these oscillations, we introduce first-order regularization into the constrained optimization of Eq.~5, 
\begin{equation}
    \mathrm{min}_\mathbf{P}
    \left(
    \|\mathbf{F}\mathbf{P}-\mathbf{M}\|_2^2
    + \lambda\|D\mathbf{P}\|_2^2
    \right),
    \label{eq:cross_val}
\end{equation}
where $D$ is the first-order finite-difference operator, $(D\mathbf{P})_i=P_{i+1}-P_i$, and $\lambda\geq0$ sets the regularization strength. We optimize $\lambda$ using two-fold cross-validation. The measurements and corresponding filters are divided into two equally sized subsets, each spanning the full range of measurement settings. For a candidate $\lambda$, we reconstruct $\mathbf{P}$ using one subset and evaluate the prediction error, $\|\mathbf{F}\mathbf{P} -\mathbf{M}\|_2$, on the other subset. We repeat this procedure after exchanging the subsets, and find the optimal $\lambda$ which minimizes the mean prediction error. The final distribution is reconstructed using the full dataset and the optimal $\lambda$. For the thermal and coherent states considered here, the optimized regularization parameter ranges from $\lambda = 0.4$ to $\lambda \approx 30$. In Table \ref{tab:fig_3_data}, we report the mean and variance of these regularized phonon reconstructions estimated using the optimal and primitive reconstruction methods.  





\bibliographystyle{apsrev4-1}
\bibliography{ref, ref_2}

\end{document}